\documentclass[prb,twocolumn]{revtex4}

\usepackage{float}

\usepackage{amsfonts}
\usepackage{amssymb}
\usepackage{amsmath}    
\usepackage{graphicx}   
\usepackage{verbatim}   
\usepackage{color}      
\usepackage{subfigure}  
\usepackage{hyperref}
\usepackage{upgreek}

\def\lb{\label}

\newcommand{\er}[1]{\textrm{(\ref{#1})}}

\begin{document}

\def\a{\alpha}  \def\cA{{\mathcal A}}     \def\bA{{\bf A}}  \def\mA{{\mathscr A}}
\def\b{\beta}   \def\cB{{\mathcal B}}     \def\bB{{\bf B}}  \def\mB{{\mathscr B}}
\def\g{\gamma}  \def\cC{{\mathcal C}}     \def\bC{{\bf C}}  \def\mC{{\mathscr C}}
\def\G{\Gamma}  \def\cD{{\mathcal D}}     \def\bD{{\bf D}}  \def\mD{{\mathscr D}}
\def\d{\delta}  \def\cE{{\mathcal E}}     \def\bE{{\bf E}}  \def\mE{{\mathscr E}}
\def\D{\Delta}  \def\cF{{\mathcal F}}     \def\bF{{\bf F}}  \def\mF{{\mathscr F}}
\def\c{\chi}    \def\cG{{\mathcal G}}     \def\bG{{\bf G}}  \def\mG{{\mathscr G}}
\def\z{\zeta}   \def\cH{{\mathcal H}}     \def\bH{{\bf H}}  \def\mH{{\mathscr H}}
\def\e{\eta}    \def\cI{{\mathcal I}}     \def\bI{{\bf I}}  \def\mI{{\mathscr I}}
\def\p{\psi}    \def\cJ{{\mathcal J}}     \def\bJ{{\bf J}}  \def\mJ{{\mathscr J}}
\def\vT{\Theta} \def\cK{{\mathcal K}}     \def\bK{{\bf K}}  \def\mK{{\mathscr K}}
\def\k{\kappa}  \def\cL{{\mathcal L}}     \def\bL{{\bf L}}  \def\mL{{\mathscr L}}
\def\l{\lambda} \def\cM{{\mathcal M}}     \def\bM{{\bf M}}  \def\mM{{\mathscr M}}
\def\L{\Lambda} \def\cN{{\mathcal N}}     \def\bN{{\bf N}}  \def\mN{{\mathscr N}}
\def\m{\mu}     \def\cO{{\mathcal O}}     \def\bO{{\bf O}}  \def\mO{{\mathscr O}}
\def\n{\nu}     \def\cP{{\mathcal P}}     \def\bP{{\bf P}}  \def\mP{{\mathscr P}}
\def\r{\rho}    \def\cQ{{\mathcal Q}}     \def\bQ{{\bf Q}}  \def\mQ{{\mathscr Q}}
\def\s{\sigma}  \def\cR{{\mathcal R}}     \def\bR{{\bf R}}  \def\mR{{\mathscr R}}
\def\cS{{\mathcal S}}     \def\bS{{\bf S}}  \def\mS{{\mathscr S}}
\def\t{\tau}    \def\cT{{\mathcal T}}     \def\bT{{\bf T}}  \def\mT{{\mathscr T}}
\def\f{\phi}    \def\cU{{\mathcal U}}     \def\bU{{\bf U}}  \def\mU{{\mathscr U}}
\def\F{\Phi}    \def\cV{{\mathcal V}}     \def\bV{{\bf V}}  \def\mV{{\mathscr V}}
\def\P{\Psi}    \def\cW{{\mathcal W}}     \def\bW{{\bf W}}  \def\mW{{\mathscr W}}
\def\o{\omega}  \def\cX{{\mathcal X}}     \def\bX{{\bf X}}  \def\mX{{\mathscr X}}
\def\x{\xi}     \def\cY{{\mathcal Y}}     \def\bY{{\bf Y}}  \def\mY{{\mathscr Y}}
\def\X{\Xi}     \def\cZ{{\mathcal Z}}     \def\bZ{{\bf Z}}  \def\mZ{{\mathscr Z}}
\def\O{\Omega}

\newcommand{\gA}{\mathfrak{A}}
\newcommand{\gB}{\mathfrak{B}}
\newcommand{\gC}{\mathfrak{C}}
\newcommand{\gD}{\mathfrak{D}}
\newcommand{\gE}{\mathfrak{E}}
\newcommand{\gF}{\mathfrak{F}}
\newcommand{\gG}{\mathfrak{G}}
\newcommand{\gH}{\mathfrak{H}}
\newcommand{\gI}{\mathfrak{I}}
\newcommand{\gJ}{\mathfrak{J}}
\newcommand{\gK}{\mathfrak{K}}
\newcommand{\gL}{\mathfrak{L}}
\newcommand{\gM}{\mathfrak{M}}
\newcommand{\gN}{\mathfrak{N}}
\newcommand{\gO}{\mathfrak{O}}
\newcommand{\gP}{\mathfrak{P}}
\newcommand{\gQ}{\mathfrak{Q}}
\newcommand{\gR}{\mathfrak{R}}
\newcommand{\gS}{\mathfrak{S}}
\newcommand{\gT}{\mathfrak{T}}
\newcommand{\gU}{\mathfrak{U}}
\newcommand{\gV}{\mathfrak{V}}
\newcommand{\gW}{\mathfrak{W}}
\newcommand{\gX}{\mathfrak{X}}
\newcommand{\gY}{\mathfrak{Y}}
\newcommand{\gZ}{\mathfrak{Z}}

\def\ve{\varepsilon}   \def\vt{\vartheta}    \def\vp{\varphi}    \def\vk{\varkappa}

\def\Z{{\mathbb Z}}    \def\R{{\mathbb R}}   \def\C{{\mathbb C}}
\def\K{{\mathbb K}}
\def\T{{\mathbb T}}    \def\N{{\mathbb N}}   \def\dD{{\mathbb D}}


\def\la{\leftarrow}              \def\ra{\rightarrow}            \def\Ra{\Rightarrow}
\def\ua{\uparrow}                \def\da{\downarrow}
\def\lra{\leftrightarrow}        \def\Lra{\Leftrightarrow}


\def\lt{\biggl}                  \def\rt{\biggr}
\def\ol{\overline}               \def\wt{\widetilde}
\def\no{\noindent}


\def\lan{\langle}                \def\ran{\rangle}
\def\/{\over}                    \def\iy{\infty}
\def\sm{\setminus}               \def\es{\emptyset}
\def\ss{\subset}                 \def\ts{\times}
\def\pa{\partial}                \def\os{\oplus}
\def\om{\ominus}                 \def\ev{\equiv}
\def\iint{\int\!\!\!\int}        \def\iintt{\mathop{\int\!\!\int\!\!\dots\!\!\int}\limits}
\def\el2{\ell^{\,2}}             \def\1{1\!\!1}
\def\sh{\sharp}
\def\wh{\widehat}
\def\bs{\backslash}

\def\all{\mathop{\mathrm{all}}\nolimits}
\def\Area{\mathop{\mathrm{Area}}\nolimits}
\def\arg{\mathop{\mathrm{arg}}\nolimits}
\def\const{\mathop{\mathrm{const}}\nolimits}
\def\det{\mathop{\mathrm{det}}\nolimits}
\def\diag{\mathop{\mathrm{diag}}\nolimits}
\def\diam{\mathop{\mathrm{diam}}\nolimits}
\def\dim{\mathop{\mathrm{dim}}\nolimits}
\def\dist{\mathop{\mathrm{dist}}\nolimits}
\def\Im{\mathop{\mathrm{Im}}\nolimits}
\def\Iso{\mathop{\mathrm{Iso}}\nolimits}
\def\Ker{\mathop{\mathrm{Ker}}\nolimits}
\def\Lip{\mathop{\mathrm{Lip}}\nolimits}
\def\rank{\mathop{\mathrm{rank}}\limits}
\def\Ran{\mathop{\mathrm{Ran}}\nolimits}
\def\Re{\mathop{\mathrm{Re}}\nolimits}
\def\Res{\mathop{\mathrm{Res}}\nolimits}
\def\res{\mathop{\mathrm{res}}\limits}
\def\sign{\mathop{\mathrm{sign}}\nolimits}
\def\span{\mathop{\mathrm{span}}\nolimits}
\def\supp{\mathop{\mathrm{supp}}\nolimits}
\def\Tr{\mathop{\mathrm{Tr}}\nolimits}
\def\BBox{\hspace{1mm}\vrule height6pt width5.5pt depth0pt \hspace{6pt}}
\def\where{\mathop{\mathrm{where}}\nolimits}
\def\as{\mathop{\mathrm{as}}\nolimits}


\newcommand\nh[2]{\widehat{#1}\vphantom{#1}^{(#2)}}
\def\dia{\diamond}

\def\Oplus{\bigoplus\nolimits}



\def\qqq{\qquad}
\def\qq{\quad}
\newcommand{\ca}{\begin{cases}}
\newcommand{\ac}{\end{cases}}
\newcommand{\ma}{\begin{pmatrix}}
\newcommand{\am}{\end{pmatrix}}
\renewcommand{\[}{\begin{equation}}
\renewcommand{\]}{\end{equation}}
\def\eq{\begin{equation}}
\def\qe{\end{equation}}
\def\[{\begin{equation}}
\def\bu{\bullet}

\title[{Shear surface waves in phononic crystals}]
      {Shear surface waves in phononic crystals}

\date{\today}

\author{A.A. Kutsenko, A.L. Shuvalov}
\affiliation{Universit\'{e} de Bordeaux, Institut de M\'{e}canique
et d'Ing\'{e}nierie de Bordeaux, UMR 5295, Talence 33405, France
}

\begin{abstract}
Existence of shear horizontal (SH) surface waves in 2D-periodic
phononic crystals with an asymmetric depth-dependent profile is
theoretically reported. Examples of dispersion spectra with band
gaps for subsonic and supersonic SH surface waves are demonstrated.
The link between the effective (quasistatic) speeds of the SH bulk
and surface waves is established. Calculation and analysis is based
on the integral form of projector on the subspace of evanescent
modes which means no need for their explicit finding. This new
method can be extended to the vector waves and the 3D case.
\end{abstract}

\maketitle

%
%
%
%
%

\section{Introduction}

Emergence of phononic crystals has reinforced the interest to
surface waves in periodic media and heighten the need for efficient
methods of calculating their dispersion branches. Considerable work
has been done for solid structures which are periodic along the
surface but uniform along the depth direction. The latter implies
pure exponential dependence on depth coordinate and thus facilitates
the plane-wave expansion (PWE) which acts on the surface coordinates
only. Applying PWE to the wave equation provides a formally infinite
algebraic system whose truncation enables explicit finding of the
evanescent (decreasing with the depth) modes which are then used to
satisfy another formally infinite system obtained by PWE of the
boundary condition on the free surface. The above two-step PWE
procedure was first implemented for Rayleigh waves in a periodic
structure of layers normal to the surface \cite{DMW} and then
extended to 2D phononic crystals composed of elastic \cite{TT,WHL}
and piezoelectric \cite{LWBK,WHH} rods normal to the surface.
Surface waves in such periodic structures uniform along the depth
direction were also calculated by FDTD \cite{SW,TTT} and wavelet
\cite{YW} methods.

By contrast, much fewer results for surface waves are available in
the alternative case of structures which are periodic both along the
surface and the depth coordinates, such as rods parallel to the
surface. We have found only three references reporting calculation
of surface-wave dispersion in depth-dependent phononic crystals
\cite{MR,Z-K,LHOA}, both papers using pure numerical means (namely,
the supercell approximation approach). One of the apparent
difficulties due to depth dependence is a numerically more involved
procedure of identifying of evanescent wave harmonics. Note that the
so-called extended PWE was suggested as a tool for this purpose
\cite{LABK,RSG,RGVHHGRSP}, but its application to the surface wave
problem in hand has not been envisaged.

The present paper pursues two objectives. The first is a new method
for calculating surface wave branches in depth-dependent phononic
crystals. The main point and advantage of the method is that the
dispersion equation is expressed in terms of the projector on the
subspace of evanescent modes and this projector is defined directly
from the material coefficients expanded in Fourier series in surface
coordinate(s), without a need to solve for partial modes and to
explicitly sort out the evanescent ones.
In principle, the proposed method can be used for general case of
vector waves in arbitrary periodic solid structures, but here it is
applied to shear horizontal (SH) waves. Study of SH surface waves in
2D phononic crystals is the second objective of the paper. This is
an interesting problem of its own right. It is well-known that the
SH surface (localised) waves in 1D periodically layered half-space
can or cannot exist if the layers are parallel or orthogonal to the
surface, respectively. At the same time, we are unaware of results
providing explicit evidence of SH surface waves in 2D periodic
structures. Uncoupling of SH modes implies 2D depth-dependent
structures, the case which defies PWE and was treated by the
supercell method in \cite{MR,Z-K,LHOA}; however, the surface waves
with SH polarization appear to be beyond the scope of this method.
As pointed out in \cite{MR}, the difficulty came from the fact that
the SH surface waves could occur only if the unit cell was
asymmetric but this made the supercell method "insufficient or
inappropriate". In contrast, asymmetry of periodic profile causes no
inconvenience for our method. By its means we demonstrate examples
of subsonic and supersonic dispersion branches of SH surface waves
in 2D phononic crystals.

The paper is organised as follows. The statement of the problem is
outlined in Sec. II. The method for calculating the surface-wave
spectrum is developed in Sec. III. Its application is exemplified in
Sec. IV. Properties of the SH surface waves and the generalization
to 3D case are discussed in Sec. V. Main findings are summarized in
Sec. VI.

\section{Statement of the problem}

Consider SH surface waves in a 2D periodic half-space $\{{\bf
x}=(x_1,x_2):x_2\ge0\}$ with a traction-free surface $x_2=0$. The
problem consists of the wave equation complemented by the boundary
and radiation conditions, namely
\[\lb{weq}
 \ca\pa_1(\m\pa_1 v)+\pa_2(\mu\pa_2v)=-\r\o^2 v,\\
 \pa_2v|_{x_2=0}=0,\ \ \lim\limits_{x_2\to\iy}v=0,\ac
\]
where $\pa_i\ev\pa/\pa x_i$ and the shear coefficient $\m({\bf x})$
and density $\r({\bf x})$ are ${\bf 1}$-periodic in $x_1$ and $x_2$.
(All subsequent results remain explicitly valid for rectangular
lattices and can be adjusted straightforwardly to the case of
oblique lattices.) Applying PWE in surface coordinate $x_1$, i.e.
inserting 1D Floquet condition along with the Fourier expansion
\begin{eqnarray}\lb{weq1}
 v({\bf x})=u({\bf x})e^{ik_1x_1}\ \ {\rm with}\ u\ {\rm periodic\ in\
 }x_1,\notag
\\
 h=\sum_m\wh h_m(x_2)e^{2\pi imx_1}\ \ {\rm for}\ \ h=u,\m,\r,
\end{eqnarray}
casts \er{weq} in the form
\[\lb{weq2}
 \ca(\pmb{\pa}+k_1)\pmb{\mu}(\pmb{\pa}+k_1){\bf u}-\pa_2(\pmb{\m}\pa_2 {\bf u})=\pmb{\r}\o^2{\bf
 u},\\ \pa_2{\bf u}|_{x_2=0}={\bf 0},\ \ \lim\limits_{x_2\to\iy}{\bf
 u}={\bf 0}\ac
\]
with
\begin{eqnarray}\label{matr}
 {\bf h}(x_2)=(\wh h_{n-m}(x_2))\ \ {\rm for}\ \ h=\m,\r,\notag\\
 \pmb{\pa}=2\pi(m\d_{n-m}),\ \ {\bf
 u}(x_2)=(\wh u_m(x_2)),\ \ n,m\in\Z.
\end{eqnarray}
For practical use, we assume all objects in \er{matr} to be of
finite dimension. Equation \er{weq2} can be rewritten as
\[\lb{ode}
 \ca\pmb{\eta}'={\bf Q}\pmb{\eta},\\
 \pmb{\eta}(0)=({\bf u}_0\ {\bf 0})^{\top},\ \ \lim\limits_{x_2\to\iy}\pmb{\eta}(x_2)={\bf 0},\ac
\]
where $'\ev\pa_2$ and
\[\lb{etaq}
 \pmb{\eta}=\ma {\bf u} \\ \pmb{\m} {\bf u}' \am,\ \
 {\bf Q}=\ma
  {\bf 0} & \pmb{\m}^{-1} \\
  (\pmb{\pa}+k_1)\pmb{\mu}(\pmb{\pa}+k_1)-\o^2\pmb{\r} &
 {\bf 0}
 \am.
\]
The solution to $\pmb{\eta}'={\bf Q}\pmb{\eta}$ with initial data
$\pmb{\eta}(0)$ is
\[\lb{mon}
 \pmb{\eta}(x_2)={\bf M}(x_2)\pmb{\eta}(0)\ \ {\rm with}\ \
 {\bf M}(x_2)=\wh{\int_0^{x_2}}({\bf I}+{\bf Q} dx_2),
\]
where $\wh{\int}$ is the multiplicative integral and ${\bf I}$ is
identity matrix. Introduce the monodromy matrix
\[\lb{prop}
 {\bf M}_0={\bf M}_0(\o,k_1)\ev{\bf M}(1)
\]
and let $q$ and ${\bf w}$ be its eigenvalues and eigenvectors.
Taking some ${\bf w}$ as initial data in \er{mon} defines the
Floquet mode $\pmb{\eta}(x_2)={\bf M}[x_2,0]{\bf w}=q^n{\bf
M}[x_2-n,0]{\bf w}$ (where $n=\lfloor x_2\rfloor$), which is either
propagating or increasing or decreasing at $x_2\to\iy$ depending on
the absolute value of the eigenvalue $q$ corresponding to ${\bf w}$.
In the case of propagating modes ($|q|=1$), the spectrum $\o=\o({\bf
k})$ (${\bf k}=(k_1,k_2)\in\R^2$) defined by the equation
$q(\o,k_1)=e^{ik_2}$ is called the Floquet spectrum.

\section{Projector-based method for calculating surface wave spectrum}

{\bf 1. Projectors.} Partition the eigenspace of ${\bf M}_0$ into
the folowing subspaces:
\[\lb{eig}
 \ca \cL_{\rm d}=\langle{\bf w}:\ {\bf M}_0{\bf w}=q{\bf w},\ |q|<1 \rangle,\\ 
     \cL_{\rm p}=\langle{\bf w}:\ {\bf M}_0{\bf w}=q{\bf w},\ |q|=1 \rangle,\\ 
     \cL_{\rm i}=\langle{\bf w}:\ {\bf M}_0{\bf w}=q{\bf w},\ |q|>1
     \rangle, 
 \ac
\]
where $\langle...\rangle$ means a span. Taking \er{mon} with
$\pmb{\eta}(0)$ from $\cL_{\rm d}$ or $\cL_{\rm p}$ or $\cL_{\rm i}$
leads to decreasing or propagating or increasing solution
$\pmb{\eta}(x_1)$, respectively. The projectors ${\bf P}_{\a}$ on
$\cL_{\a}$ ($\a={\rm d},{\rm p},{\rm i}$), i.e.
\[\lb{defpr}
 {\bf P}_{\a}{\bf w}=\ca{\bf w}, & {\bf w}\in\cL_{\a}, \\ {\bf 0}, & {\bf w}\not\in\cL_{\a},\ac
\]
can be defined by the formulas
\[\lb{proj}
 \ca {\bf P}_{\rm d}=\frac1{2\pi i}\int\limits_{|z|=1-0}(z{\bf I}-{\bf M}_0)^{-1}dz, \\
     {\bf P}_{\rm p}=\frac1{2\pi i}(\int\limits_{|z|=1+0}-\int\limits_{|z|=1-0})(z{\bf I}-{\bf M}_0)^{-1}dz, \\
     {\bf P}_{\rm i}={\bf I}-\frac1{2\pi i}\int\limits_{|z|=1+0}(z{\bf I}-{\bf M}_0)^{-1}dz.
 \ac
\]
By definition
\[\lb{projs}
 {\bf P}_{\rm d}+{\bf P}_{\rm p}+{\bf P}_{\rm i}={\bf
I}.
\]
Note that both ${\bf P}_{\rm p}$ and ${\bf P}_{\rm i}$ can be
expressed in terms of ${\bf P}_{\rm d}$, see \er{pproj1}.

{\bf 2. Dispersion equation.} The surface wave problem \er{ode} is
equivalent to any one of the following conditions
\begin{eqnarray}\lb{surfcon1}
 \exists {\bf u}_0&\ne&{\bf 0}:\ \ \ma{\bf u}_0 \\ {\bf 0}\am\in\cL_{\rm
 d}\ \Leftrightarrow \\
 \lb{surfcon2}
 \exists {\bf u}_0&\ne&{\bf 0}:\ \ {\bf P}_{\rm d}\ma{\bf u}_0 \\ {\bf 0}\am=\ma{\bf u}_0 \\ {\bf
 0}\am\ \Leftrightarrow\\
 \lb{surfcon3}
 \exists{\bf u}_0&\ne&{\bf 0}:\ {\bf P}_{{\rm d}3}{\bf u}_0={\bf 0},\ \ {\bf P}_{{\rm
 d}1}{\bf u}_0={\bf u}_0,
\end{eqnarray}
where
\[\lb{pdel}
 {\bf P}_{\rm d}=\ma {\bf P}_{{\rm d}1} & {\bf P}_{{\rm d}2} \\
 {\bf P}_{{\rm d}3} & {\bf P}_{{\rm d}4} \am.
\]
According to \er{surfcon3}, the surface wave spectrum $\o_{\rm
saw}(k_1)$ can be defined by the dispersion equation
\[\lb{disp}
 D_{\rm saw}(\o,k_1)\ev\det(({\bf P}_{{\rm d}1}-{\bf I})^*({\bf P}_{{\rm d}1}-{\bf I})+
 {\bf P}_{{\rm d}3}^*{\bf P}_{{\rm d}3})=0,
\]
where $^*$ means Hermitian conjugation.

{\bf 3. Projection of the Floquet spectrum on the plane $(\o,k_1)$.}
Recall that the Floquet  spectrum $\o({\bf k})$ can be determined by
the equation
\[\lb{floq1}
 \exists{\bf w}\ne{\bf 0}:\ \ {\bf
 M}_0(\o,k_1){\bf w}=e^{ik_2}{\bf w}.
\]
Introduce the multiplicity of the projection of the Floquet spectrum
on the plane $(\o,k_1)$
\[\lb{multpr}
 N_{\rm p}(\o,k_1)=\#\{k_2\in\R:\ \o({\bf k})=\o\},
\]
which indicates the number of propagating modes. By \er{eig},
\er{defpr} and \er{floq1}, it can be evaluated as
\[\lb{nofpm}
 N_{\rm p}(\o,k_1)=\dim\cL_{\rm p}={\rm Trace}\ {\bf P}_{\rm p}.
\]
By definition, $N_{\rm p}$ is a piecewise constant function with
integer even values. Denote the boundaries of areas where $N_{\rm
p}$ takes the same value by $\o_{\rm tr}(k_1)$ and call them {\it
transonic} curves. These curves coincide with the projection of
local extrema of Floquet branches $\o({\bf k})$ on the plane
$(\o,k_1)$. The areas of $(\o,k_1)$-plane where $N_{\rm
p}\ne0(\Leftrightarrow{\bf P}_{\rm p}\ne{\bf 0})$ and $N_{\rm
p}=0(\Leftrightarrow{\bf P}_{\rm p}={\bf 0})$ will be referred to as
{\it propagative} and {\it non-propagative} domains, respectively.
The whole subsonic range $\{(\o,k_1):\ \o<\o_{{\rm
tr},1}(k_1)=\min_{k_2}\o({\bf k})\}$, i.e. the part of the plane
$(\o,k_1)$ below the minimal frequency of the Floquet spectrum
$\o({\bf k})$, is always non-propagative by definition. Owing to
existence of spectral gaps in $\o({\bf k})$, the non-propagative
domains may also arise in the supersonic range, i.e. above $\o_{{\rm
tr},1}(k_1)$.

{\bf 4. Refined procedure in the non-propagative domains.} Consider
a non-propagative domain $N_{\rm p}(\o,k_1)=0$. According to
\er{surfcon3} and the identity ${\bf P}_{\rm d}+{\bf P}_{\rm i}={\bf
I}$ (see \er{projs}), any root $(\omega,k_{1}) $ of the equation
\[\lb{acon2}
 D_{{\rm d}3}(\o,k_1)\ev\det{\bf P}_{{\rm d}3}=0
\]
corresponds to a surface-wave solution $\o_{\rm saw}(k_1)$ (see
\er{surfcon1}) or a nonphysical solution $\o_{\rm i}(k_1)$ which
contains of increasing modes (i.e. $\ma {\bf u}_0 & {\bf 0}
\am^{\top}\in\cL_{\rm i}$ instead of \er{surfcon1}). Note that
$D_{{\rm d}3}(\o,k_1)$ is a real function for real arguments, since
${\bf P}_{{\rm d}3}$ is a self-adjoint matrix at $N_{\rm p}=0$, see
\er{pdel1}. Unlike non-negative $D_{\rm saw}$ in \er{disp}, the
function $D_{{\rm d}3}$ generally changes sign at its zeroes. Thus
seeking the surface waves in non-propagative domains, it is
convenient to use \er{acon2} alongside \er{disp}: the former
verifies that a "numerical zero" is not a deep but nonzero minimum
and the latter checks out whether this zero defines a surface wave
rather than a non-physical wave. Note that surface waves can also be
identified by using \er{acon2} along with the condition ${\bf
P}_{{\rm d}1}{\bf u}_0\ne{\bf 0}$, where ${\bf u}_0$ is defined by
${\bf P}_{{\rm d}3}{\bf u}_0={\bf 0}$.

In the propagative domains $N_{\rm p}(\o,k_1)\ne0$, due to
$\rank{\bf P}_{\rm d}=\rank{\bf P}_{\rm i}$ and ${\bf P}_{\rm
p}\ne{\bf 0}$ it follows from \er{projs} that $\rank{\bf P}_{{\rm
d}3}\le\rank{\bf P}_{\rm d}<\frac12\rank{\bf I}=\dim{\bf P}_{{\rm
d}3}$. Therefore \er{acon2} is an identity in propagative domains
and so it cannot be used for defining $\o_{\rm saw}(k_1)$ in
propagative domains.

{\bf 5. Calculation of projectors.} Equation \er{proj} defines the
projector $\mathbf{P}_{\mathrm{d}}$ as an integral of the resolvent
$\mathbf{R} _{z0}=( z{\bf I}-{\bf M}_{0}) ^{-1}.$ One way to obtain
$\mathbf{R} _{z0}$ is to compute $\mathbf{M}_{0}$ defined by
\er{mon}. However, computing $\mathbf{M}_{0}$ of large algebraic
dimension $2d\times 2d$ (which means taking into account many
members of the Fourier series, see \er{matr}) is numerically
troublesome because some of its components grow exponentially as $d$
increases. On the other hand, components of $\mathbf{R}_{z0}$ can
grow in general only as fast as linearly in $d$. Therefore it is
numerically adavantageous to calculate $\mathbf{R}_{z0}$ directly
rather than via $ \mathbf{M}_{0}$.

Denote
\[\lb{resolvent}
\mathbf{R}_{\alpha }(x_2) =( \alpha{\bf I}-{\bf M}(x_2)) ^{-1}
\]
where $\alpha $ is some fixed constant which does not belong to the
spectrum of $\mathbf{M}$. As a solution to \er{ode}, $\mathbf{M}$
satisfies the linear differential equation with initial data
\[\lb{diffres}
\ca \mathbf{M}'(x_{2})=\mathbf{Q}( x_{2}) \mathbf{M}(x_2),
\\
\mathbf{M}(0)=\mathbf{I}, \ac
\]
from which it follows that $\mathbf{R}_{\alpha }$ satisfies the
Ricatti equation
\[\lb{ricatti}
 \ca{\bf R}_{\a}'(x_2)={\bf R}_{\a}(x_2){\bf Q}(x_2)(\a{\bf R}_{\a}(x_2)-{\bf
 I}),\\ {\bf R}_{\a}(0)=(\a-1)^{-1}{\bf I}.\ac
\]
Numerical integration of \er{ricatti} provides the value
$\mathbf{R}_{\alpha 0}={\bf R}_{\a}(1)$ for a given $\alpha$. Once
it is found, the identity
\[\lb{resid}
 {\bf R}_{z0}={\bf R}_{\a0}({\bf I}+(z-\a){\bf R}_{\a0})^{-1}
\]
yields $\mathbf{R}_{z0}$ with varying $z$ as required in \er{proj}.
Thus combining \er{proj} and \er{resid} defines the projector
$\mathbf{P}_{\mathrm{d}}$ in the following specialized form
\[\lb{pdres}
 {\bf P}_{\rm d}=\frac{{\bf R}_{\a0}}{2\pi}\int_0^{2\pi}re^{i\vp}({\bf I}+(re^{i\vp}-\a){\bf
 R}_{\a0})^{-1}d\vp
\]
with $r<1$ sufficiently close to $1$. Two other projectors
$\mathbf{P}_{\mathrm{i}}$ and $\mathbf{P}_{ \mathrm{p}}$ are
expressed via $\mathbf{P}_{\mathrm{d}},$ see \er{pproj1}.

In numerical implementation, we used the fourth order Runge-Kutta
method for calculating $\mathbf{R}_{\alpha 0}$ from \er{ricatti}
where $\alpha \in\C $ has being chosen randomly anew for each next
calculation point $( \omega ,k_{1})$. It is recommended to avoid
taking $\alpha $ near the real axis or the unit circle (we used
$\alpha \in[ -6,-3] \times [ 3i,6i] $). The Chebyshev method was
employed for evaluating \er{pdres} where we took $r=0.99$. Dealing
with Eqs. \er{disp} and \er{acon2} it is convenient to use
'normalised' functions
\begin{eqnarray}\lb{normalised}
D_{\mathrm{saw}}(\omega ,k_{1})&=&\lambda_{\min
}(\omega,k_{1}),\notag\\
D_{\mathrm{d}3}(\omega,k_{1})&=&\det(\mathbf{P}_{\mathrm{d}3}^{-1}(
0,k_{1})\mathbf{P}_{\mathrm{d}3}(\omega,k_{1})),
\end{eqnarray}
where $\lambda _{\min }$ is the minimal eigenvalue of the
self-adjoint non-negative operator whose determinant is taken in
\er{disp}.

\section{Examples}

\textbf{1. Stiff cylinders in a soft matrix.} Assume an epoxy matrix
with a periodic structure of steel cylindrical bars parallel to the
free surface $x_{2}=0$. The material constants are $\rho =1.14\ {\rm
g/cm}^{3}$,\ $\mu =1.48\ {\rm GPa}$ for epoxy and $\rho =7.8\ {\rm
g/cm}^{3}$, $\mu =80\ {\rm GPa}$ for steel. Let the steel bars form
a rectangular lattice with the horizontal and vertical periods
$a_{1}:a_{2}=1:2$, and the radius of bars be $0.45a_{1}$ (in the
following, $a_{1}\ev1$ and $k_{1}a_{1}\ev k_{1}$, $\omega a_{1}\ev
\omega$ in units ${\rm mm/\upmu s}$). For obtaining SH surface
waves, it is necessary that the unit cell is asymmetric about the
horizontal midplane (see Sec. V.1). We shall consider two
interrelated types of structures which are {\it reciprocal} to each
other in the sense that their unit cells pass into one another by
means of reflection about the horizontal midplane, see Fig. 1. Note
that the unit cell of the structure (a) is 'faster' on the top than
on the bottom, hence vice versa for the structure (b). Figure 2
demonstrates that the two reciprocal structures are characterized at
fixed $(\omega,k_{1})$ by the same number $N_{\mathrm{p}}$ of
propagating modes, whereas the frequency $\omega _{\mathrm{saw}}$ of
surface-wave solutions for the configuration (b) is the frequency
$\omega _{\mathrm{i}}$ of non-physical solutions for the
configuration (a). A general proof for this feature is given in Sec.
V.2.

The surface-wave dispersion branches $\omega _{\mathrm{saw}}(
k_{1})$ for the structure (b) are shown in Fig. 3. For this
calculation, we used $d=17$ terms of the Fourier series \er{weq1}.
Numerical data clearly show that there exists a subsonic branch
$\omega_{\mathrm{saw},1}$ below the first transonic curve
$\omega_{\mathrm{tr},1}$, though their relative difference  is only
of the order of $1\%$. Subsonic and supersonic surface waves
occurring in the non-propagative domains ($N_{\mathrm{p}}=0$) are
defined by common zeros of Eqs. \er{disp} and \er{acon2}. In
propagative domains ($N_{\mathrm{p}}\ne0$), a surface wave is
defined only by \er{disp} and is therefore associated, in the
numerical context, with a minimum which tends to zero with growing
number $d$ of terms of Fourier series. The surface-wave branches
shown within the upper propagative domain in Fig. 3 yield the value
of $D_{\mathrm{saw}}=\lambda _{\min }$ (see \er{normalised}) of
about $10^{-3}$.

\begin{figure}[h] \centering
\includegraphics[width=0.7\linewidth]{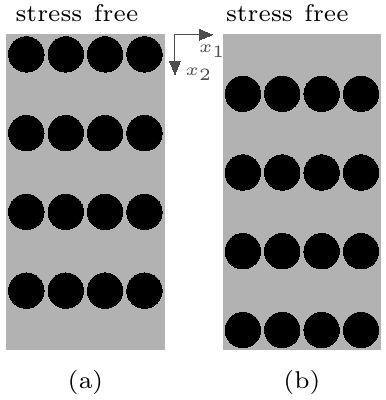}
{\caption{Reciprocal configurations of a periodic structure of steel
cylinders in epoxy matrix.}} \label{fig1}
\end{figure}

\begin{figure}[h] \centering
\begin{minipage}[h]{0.8\linewidth}
\centering\includegraphics[width=0.99\linewidth]{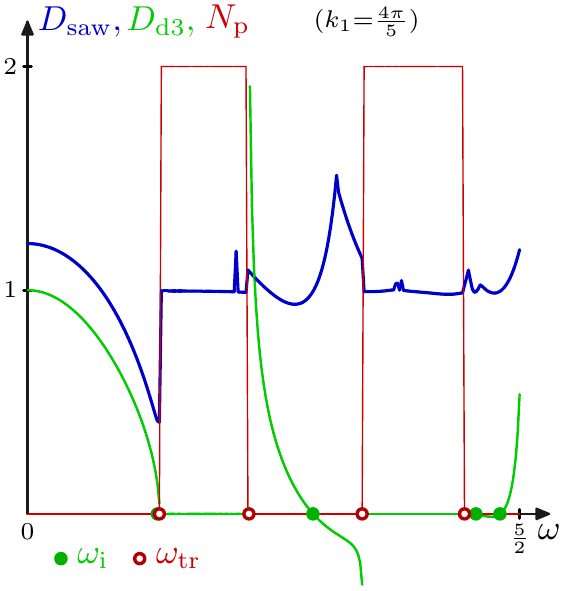} \\
(a)
\end{minipage}
\vfil
\begin{minipage}[h]{0.8\linewidth}
\centering\includegraphics[width=0.99\linewidth]{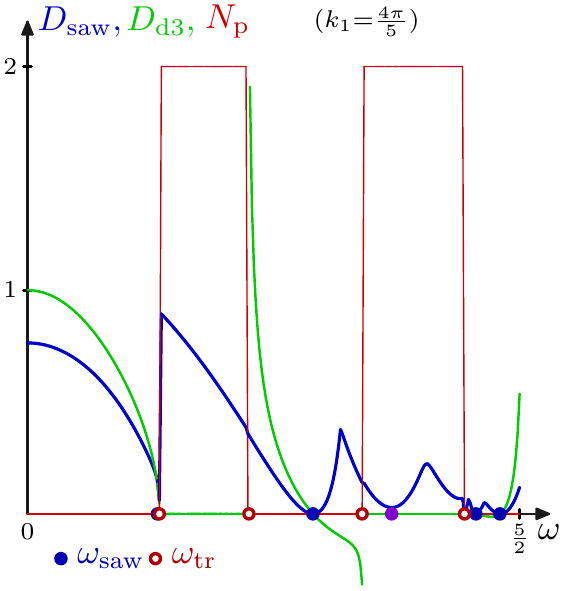}
\\ (b)
\end{minipage}
\caption{(color online) The function $N_{\mathrm{p}}(\omega) $ (see
\er{nofpm}) and the normalized functions $D_{\mathrm{saw}}(\omega)$
and $\ D_{ \mathrm{d}3}(\omega)$ (see \er{normalised}) calculated at
fixed $k_{1}=4\pi /5$ for the reciprocal structures (a) and (b) of
Fig. 1, respectively. The curves $D_{\mathrm{d}3}$ and
$N_{\mathrm{p}}$ do not change on passing from one configuration to
another, see \er{recippr}. Zeros of $D_{\mathrm{d}3}$ at
$N_{\mathrm{p}}=0$ are surface-wave solutions $\omega _{
\mathrm{saw}}$ when they coincide with zeros of $D_{\mathrm{saw}}$,
otherwise they are non-physical solutions $\omega _{\mathrm{i}}$.}
\label{fig2}
\end{figure}

\begin{figure}[h] \centering
\includegraphics[width=0.9\linewidth]{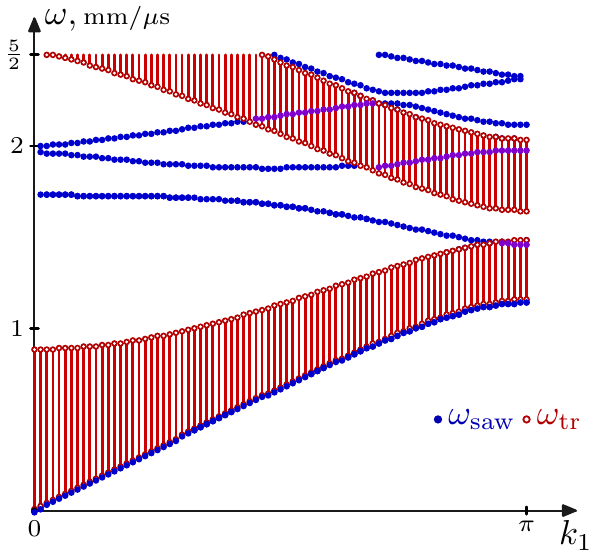}
{\caption{Surface-wave dispersion branches $\omega
_{\mathrm{saw}}(k_{1})$ and transonic curves $\omega
_{\mathrm{tr}}(k_{1}) $ for the structure displayed in Fig. 1b.
Blank and hatched areas are the non-propagative and propagative
domains, respectively.}} \label{fig3}
\end{figure}

{\bf 2. Perturbation of 1D-periodic structure.}  In order to
illuminate formation of surface-wave spectrum, let us think of a
vertically periodic stack of equidistant epoxy and iron layers (Fig.
4a) and then assume that each epoxy layer contains narrow lead
plates embedded periodically along the horizontal direction (Fig.
4b). The material constants are $\rho=116\ {\rm g/cm }^{3}$, $\mu
=7.88\ {\rm GPa}$ for Fe and $\rho=14.9\ {\rm g/cm}^{3}$, $\mu
=11.6\ {\rm GPa}$ for Pb. Taking the ratio of the horizontal to
vertical periods $a_{1}:a_{2}=1:3$ for both cases and applying the
calculation procedure of Sec. III leads to the spectra shown in
Figs. 5a and 5b, respectively. It is clearly seen that introducing
horizontal periodicity creates band gaps at the edge and inside of
the Brillouin zone.

\begin{figure}[h] \centering
\includegraphics[width=0.7\linewidth]{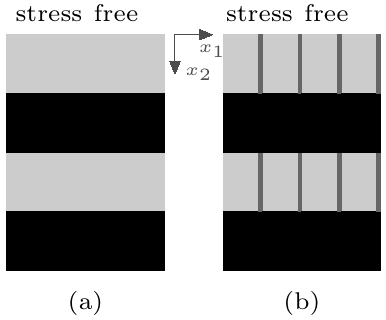}
{\caption{Periodic structure of epoxy and iron layers (a) and a
similar structure but with lead baffles in epoxy (b). Lead to epoxy
volume fraction is $1:9$.}} \label{fig4}
\end{figure}

\begin{figure}[h] \centering
\begin{minipage}[h]{0.9\linewidth}
\centering\includegraphics[width=0.99\linewidth]{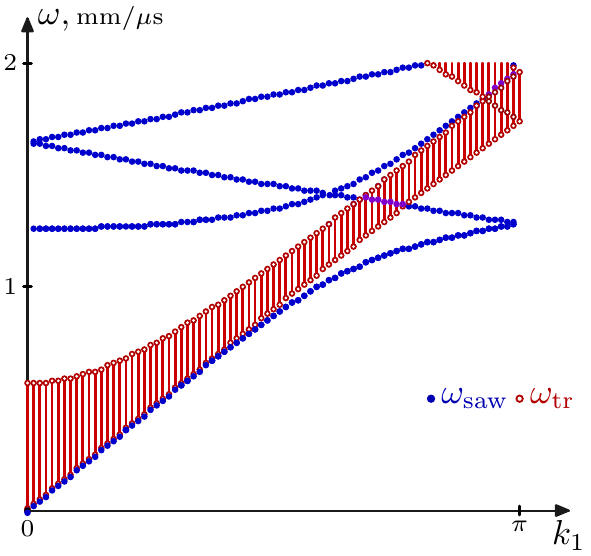} \\
(a)
\end{minipage}
\vfil
\begin{minipage}[h]{0.9\linewidth}
\centering\includegraphics[width=0.99\linewidth]{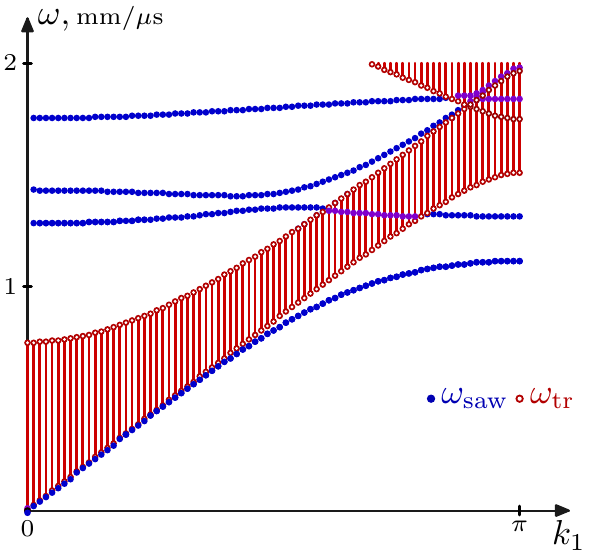}
\\ (b)
\end{minipage}
\caption{Surface-wave dispersion spectra $\omega _{\mathrm{saw}}(
k_{1}) $ (a) and (b) for the structures displayed in Fig. 4a and 4b,
respectively. Notations are the same as in Fig. 3.} \label{fig5}
\end{figure}

\section{Discussion}

{\bf 1. Reciprocity property for SH surface waves.} Given periodic
functions $\mu(\mathbf{x})$ and $\rho(\mathbf{x})$ of $ \mathbf{x}=(
x_{1},x_{2})\in \R^2$, let us mentally cut the space into two halves
by the plane $x_{2}=0$ and turn the half-space $x_{2}\leq 0$ upside
down. Thus we obtain two models of a half-space $x_{2}\geq 0$: one
with the profile $\mu(\mathbf{x})$, $\rho(\mathbf{x})$ and another
with the reciprocal profile $\widetilde{\mu}(\mathbf{x})
=\mu(x_{1},-x_{2})$, $\widetilde{\rho }(\mathbf{x}) =\rho(
x_{1},-x_{2})$. Consider the relation between the properties for a
direct and reciprocal profiles (the objects constructed from
$\widetilde{\mu }$, $\widetilde{\rho }$ will be labeled by a tilde).

By \er{mon} and the definition of multiplicative integral,
\[\lb{sprop}
\begin{array}{c}\mathbf{M}_{0}^{-1}=\lim\limits_{N\to \infty
}\prod\limits_{j=N}^{1}\lt(
\mathbf{I}-\frac1N\mathbf{Q}\left(\frac{N-j}{N}\right)\rt)
\\
=\lim\limits_{N\rightarrow \infty }\prod\limits_{j=N}^{1}\lt(
\mathbf{I}-\frac1N\widetilde{\mathbf{Q}}\left(
\frac{j}{N}\right)\rt)=\mathbf{S}\widetilde{\mathbf{M}}_{0}\mathbf{S}
\\
{\rm with}\ \ {\bf S}={\bf S}^{-1}=\ma
\mathbf{I} & \mathbf{0} \\
\mathbf{0} & -\mathbf{I}\am,
\end{array}
\]
where we used that
$\mathbf{S}\widetilde{\mathbf{Q}}\mathbf{S=-}\widetilde{\mathbf{Q}}$
since $\widetilde{\mathbf{Q}}$ has zero diagonal blocks, see
\er{etaq}. According to \er{sprop}, the eigen-subspaces and
projectors for direct and reciprocal profiles are related as
follows:
\[\lb{pp001}
 \ca
 \wt\cL_{\rm d}={\bf S}\cL_{\rm i}, & \wt{\bf P}_{\rm d}={\bf S}{\bf P}_{\rm i}{\bf S},\\
 \wt\cL_{\rm i}={\bf S}\cL_{\rm d}, & \wt{\bf P}_{\rm i}={\bf S}{\bf P}_{\rm d}{\bf S},\\
 \wt\cL_{\rm p}={\bf S}\cL_{\rm p}, & \wt{\bf P}_{\rm p}={\bf S}{\bf P}_{\rm p}{\bf S}.\ac
\]
Using a similar to \er{pdel} notation for the blocks of ${\bf
P}_{\rm i}$, introduce the equation
\[\lb{dispinv}
 D_{\rm i}(\o,k_1)\ev\det(({\bf P}_{{\rm i}1}-{\bf I})^*({\bf P}_{{\rm i}1}-{\bf I})+
 {\bf P}_{{\rm i}3}^*{\bf P}_{{\rm i}3})=0,
\]
whose solutions $\o_{\rm i}(k_1)$ describe non-physical waves which
satisfy the stress-free condition but consist of increasing modes.
From \er{pp001}, with reference to \er{disp}, \er{dispinv} and
\er{nofpm},
\[\lb{recippr}
D_{\mathrm{saw}}=\widetilde{D}_{\mathrm{i}},\
D_{\mathrm{i}}=\widetilde{D}_{\mathrm{saw}},\ D_{{\rm d}3}=\wt
D_{{\rm d}3},\ N_{\mathrm{p}}=\widetilde{N}_{\mathrm{p}}.
\]
It is thus proved that (i) the surface wave solution $\omega
_{\mathrm{saw}}( k_{1}) $ for a direct profile $\mu $, $\rho$  is at
the same time a "non-physical" solution $\widetilde{\omega
}_{\mathrm{i}}( k_{1}) $ for a reciprocal profile $\widetilde{\mu
}$, $\widetilde{\rho }$ and vice versa, and that (ii) the number of
propagating modes for direct and reciprocal profile is the same.
Also we obtain from \er{pp001} that (iii) {\it there is no SH
surface waves if the profile $\m$, $\r$ is symmetric in the depth
direction, i.e. if $\m=\wt\m$, $\r=\wt\r$}. Indeed, assume the
opposite: there exists a surface wave in the case of a symmetric
profile. Then using \er{surfcon1}, \er{pp001} and
$\m,\r=\wt\m,\wt\r$ yields
\[\lb{even}
 \cL_{\rm d}\ni\ma{\bf u}_0 \\ {\bf 0}\am={\bf S}\ma{\bf u}_0 \\ {\bf
 0}\am\in{\bf S}\cL_{\rm d}=\wt\cL_{\rm i}=\cL_{\rm i},
\]
which is in contradiction with $\cL_{\rm d}\cap\cL_{\rm i}=\{{\bf
0}\}$.

{\bf 2. Effective (quasistatic) speed of the fundamental SH surface wave. }%
The surface-wave spectrum $\omega _{\mathrm{saw}}\left( k_{1}\right)
$ may or may not contain the so-called fundamental branch which
starts at zero $\omega$ and $k_{1}$. Suppose that it does, i.e. that
the first (lowest) branch $\omega _{\mathrm{saw},1}\left(
k_{1}\right) $ is a fundamental branch. Denote
\begin{eqnarray}\label{ceff}
c_{\mathrm{saw}} &=&\lim\limits_{k_{1}\rightarrow 0}\frac{\omega _{\mathrm{%
saw},1}( k_{1}) }{k_{1}},\  \notag\\
c( \pmb{\kappa })  &=&\lim_{|\mathbf{k}|
\rightarrow 0}\frac{\omega_{1}(\mathbf{k}) }{|\mathbf{%
k}|}\ \lt( \pmb{\kappa }  =\frac{\mathbf{k}}{%
|\mathbf{k}|}\rt),\  \\
c_{\mathrm{tr}} &=&\lim_{k_{1}\rightarrow 0}\frac{\omega _{\mathrm{tr}%
,1}( k_{1}) }{k_{1}}\notag
\end{eqnarray}%
where $c_{\mathrm{saw}}$ is the effective speed of the fundamental
SH surface wave, $c\left( \pmb{\kappa }\right) $ is the effective
speed of bulk Floquet modes ($\omega _{1}(\mathbf{k}) $ is the
lowest sheet of the Floquet spectrum), and $c_{\mathrm{tr}}$ is the
onset slope of the first transonic curve $\omega _{\mathrm{tr},1}(
k_{1})$. It can be proved that
\[
c_{\mathrm{saw}}=c_{\mathrm{tr}}.
\]%
On the other hand, from \er{ceff} and the definition $\omega
_{\mathrm{tr},1}( k_{1}) =\min_{k_{2}}\omega_1(\mathbf{k})$ we
deduce that $c^2_{\rm
tr}=\min_{\frac{\k_2}{\k_1}}\k_1^{-2}c^2(\pmb{\k})$. Recalling that
a squared $c(\pmb{\kappa })$ is a quadratic form of $\pmb{\kappa}$,
it follows that
\[
c_{\mathrm{tr}}^{2}=\frac{1}{C_{22}}\det \left( C_{ij}\right),\
\mathrm{where}\ c^{2}(\pmb{\kappa})=\kappa _{i}C_{ij}\kappa _{j}\ \
(i,j=1,2).
\]
Explicit expressions for the components of the matrix $\left(
C_{ij}\right)$ can be found in Sec. IIC of \cite{KSN}. Note that
this matrix is diagonal for the structures displayed in Fig. 1 and
non-diagonal for the structure displayed in Fig. 4b.

{\bf 3. Algebraic symmetries.} Recall that \er{etaq} and \er{mon}
imply the standard identities
\[\lb{qp1}
 {\bf Q}^{*}=-{\bf T}^{-1}{\bf Q}{\bf T},\ \ {\bf M}^{-1}={\bf T}^{-1}{\bf
 M}^{*}{\bf T}
\]
where
\[\lb{TT}
 {\bf T}=-{\bf
 T}^{*}=-{\bf T}^{-1}=\ma {\bf 0} & {\bf I} \\ -{\bf I} & {\bf 0}\am.
\]
Using \er{qp1} with \er{proj} gives
\begin{eqnarray}\lb{idp1}
 {\bf P}_{\rm d}^*&=&\frac{-1}{2\pi i}\int\limits_{|z|=1-0}({z}^*{\bf I}-{\bf
 M}_0^*)^{-1}d{z}^*\notag
\\
 &=&\frac{1}{2\pi i}\int\limits_{|z|=1-0}({z}{\bf I}-{\bf
 M}_0^*)^{-1}d{z}\notag
\\
 &=&{\bf T}\frac{1}{2\pi i}\int\limits_{|z|=1-0}({z}{\bf I}-{\bf T}{\bf
 M}_0^*{\bf T}^*)^{-1}d{z}{\bf T}^*\notag
\\
 &=&{\bf T}\frac{1}{2\pi i}\int\limits_{|z|=1-0}({z}{\bf I}-{\bf
 M}_0^{-1})^{-1}d{z}{\bf T}^*
 ={\bf T}^{-1}{\bf P}_{\rm i}{\bf
 T}.\notag
\end{eqnarray}
Thus, with reference to \er{projs}, knowing the projector
$\mathbf{P}_{\mathrm{d}}$ yields the two other projectors as follows
\begin{eqnarray}\lb{pproj1}
 {\bf P}_{\rm i}&=&{\bf T}^{-1}{\bf P}_{\rm d}^*{\bf T},\notag
 \\
 {\bf P}_{\rm p}&=&{\bf I}-{\bf P}_{\rm d}-{\bf T}^{-1}{\bf P}^*_{\rm d}{\bf T}
 \ \ (={\bf T}^{-1}{\bf P}_{\rm p}^*{\bf T}).
\end{eqnarray}
In the non-propagative domains $N_{\rm p}=0(\Leftrightarrow
\mathbf{P}_{\rm p}=0)$, Eq. \er{projs} reduces to
$\mathbf{P}_{\mathrm{d}}+{\bf T}^{-1}{\bf P}_{
\mathrm{d}}^{*}\mathbf{T}=\mathbf{I}$ and so the blocks \er{pdel} of
$ \mathbf{P}_{\mathrm{d}}$ satisfy
\[\lb{pdel1}
 {\bf P}_{{\rm d}2,{\rm d}3}^*={\bf P}_{{\rm
 d}2,{\rm d}3},\ \ {\bf P}_{{\rm d}1}+{\bf P}_{{\rm d}4}^*={\bf I}.
\]

By \er{qp1}, the $x_{2}$-component of energy flux averaged over
$x_{1}$ and time, $\mathcal{P}(x_{2}) =-\frac{i\omega }{4}\pmb{\eta
}^{*}\mathbf{T}\pmb{\eta }$, is non-zero or zero if, respectively,
$\pmb{\eta }(x_{2}) =\mathbf{M}(x_{2})\mathbf{w}$ is a propagating
or non-propagating Floquet mode, as defined in Sec. II. Equal number
of positive and negative values of $\mathcal{P}$, i.e. of forward
and backward propagating modes, is due to zero signature of
$\mathbf{T}$.

In conclusion, note that the definition of $\pmb{\eta }$ in
\er{etaq} differs from another conventional form which involves an
additional factor '$i$' and hence leads to the similar identities
for $\mathbf{Q}\ $and $\mathbf{M}$ but with $\mathbf{T}$ having both
off-diagonal blocks equal to $\mathbf{I}$ (unlike \er{TT}).

{\bf 4. Spectral decomposition in the non-propagative domains.} The
integral definition \er{proj} of the projectors underlying the
present method on the whole and the above identities in particular
is independent of accidental degeneracies of eigenvalues $q$ of
$\mathbf{M}_{0}$. Let us assume a generic case where all $q$ are
distinct and hence $ \mathbf{M}_{0}$ possesses a full set of
eigenvectors $\mathbf{w}$. Then besides \er{proj} projectors satisfy
the spectral decomposition. Consider a non-propagative domain (all
$|q|\ne1$). Apply the normalization
$\mathbf{w}_{\mathrm{d}}^{*}\mathbf{Tw}_{\mathrm{i}}=1$ where $
\mathbf{w}_{\mathrm{d}}$ and $\mathbf{w}_{\mathrm{i}}$ correspond to
$|q|<1$ and $|q|>1,$ i.e. to decreasing and increasing Floquet
modes. Define the matrix $\mathbf{W}$
whose columns are sets of normalized $\mathbf{w}_{\mathrm{d}}=\left( \mathbf{%
a}_{\mathrm{d}}\mathbf{\ b}_{\mathrm{d}}\right) ^{\top}$ and $\mathbf{w%
}_{\mathrm{i}}=\left( \mathbf{a}_{\mathrm{i}}\mathbf{\ b}_{\mathrm{i}%
}\right) ^{\top},$ namely,
\[
\mathbf{W}=\ma
\mathbf{A}_{\mathrm{d}} & \mathbf{A}_{\mathrm{i}} \\
\mathbf{B}_{\mathrm{d}} & \mathbf{B}_{\mathrm{i}} \am
\]
where blocks $\mathbf{A}_{\mathrm{d},\mathrm{i}}$ and $\mathbf{B}_{\mathrm{d}%
,\mathrm{i}}$ consist of $\mathbf{a}_{\mathrm{d},\mathrm{i}}$ and
$\mathbf{b}_{\mathrm{d},\mathrm{i}}$. By \er{qp1}, ${\bf W}{\bf
T}^*{\bf W}^{*}{\bf T}={\bf I}$ and hence
\begin{eqnarray}
\mathbf{P}_{\mathrm{d}}=\ma
\mathbf{A}_{\mathrm{d}} & \mathbf{0} \\
\mathbf{B}_{\mathrm{d}} & \mathbf{0} \am
\ma
\mathbf{B}_{\mathrm{i}}^{*} & -\mathbf{A}_{\mathrm{i}}^{*} \\
\mathbf{0} & \mathbf{0}\am =\ma
\mathbf{A}_{\mathrm{d}}\mathbf{B}_{\mathrm{i}}^{*} & -\mathbf{A}_{\mathrm{d}}%
\mathbf{A}_{\mathrm{i}}^{*} \\
\mathbf{B}_{\mathrm{d}}\mathbf{B}_{\mathrm{i}}^{*} & -\mathbf{B}_{\mathrm{d}}%
\mathbf{A}_{\mathrm{i}}^{*}%
\am,  \\
\mathbf{P}_{\mathrm{i}}=\ma
\mathbf{0} & \mathbf{A}_{\mathrm{i}} \\
\mathbf{0} & \mathbf{B}_{\mathrm{i}} \am
\ma
\mathbf{0} & \mathbf{0} \\
-\mathbf{B}_{\mathrm{d}}^{*} & \mathbf{A}_{\mathrm{d}}^{*}\am =\ma
-{\bf A}_{\rm i}{\bf B}_{\rm d}^* & {\bf A}_{\rm i}{\bf
A}_{\rm d}^* \\
-{\bf B}_{\rm i}{\bf B}_{\rm d}^* & {\bf B}_{\rm i}{\bf A}_{\rm
d}^*\am.
\end{eqnarray}
Note that the structure of the left off-diagonal block ${\bf
P}_{{\rm d}3}$ corroborates with the interpretation of Eq.
\er{acon2} given in Sec. III.4.

{\bf 5.  General 3D case and the depth-independent case.} The
projector-based method described in Sec. III can be extended to
vector surface waves (Rayleigh waves) and to 3D periodic materials
by means of a single replacement of the SH matrix ${\bf Q}$ by its
general form
\begin{eqnarray}\lb{3dq}
 {\bf Q}=\ma -\cC^{-1}\cA_1 & \cC^{-1} \\
 -\r\o^2-\cA_2-\cA_1^*\cC^{-1}\cA_1 & \cA_1^*\cC^{-1} \am\notag,\\
 \cC=(c_{i1k1}),\ \ \cA_1=(c_{i1kl}(\pa_l+ik_l)),\notag\\
 \cA_2=((\partial_j+ik_j)c_{ijkl}(\partial_l+ik_l)),\ \ j,l=1,2,
\end{eqnarray}
where $c_{ijkl}({\bf x})$ are the stiffness coefficients. For
brevity, \er{3dq} is written in the ${\bf x}$-space and is subject
to Fourier expansion in the surface coordinate(s). Note that
\er{3dq} with $k_{1,2}=0$ reduces to (5.28) of \cite{KSN}.

If $c_{ijkl}$ are uniform in the depth direction $x_3$, then the
definition of projectors ${\bf P}$ can be simplified due to
replacing ${\bf M}_0$ in \er{proj} by
\[\lb{3dmon}
 {\bf L}=({\bf Q}+\a{\bf I})({\bf Q}-\a^*{\bf I})^{-1}
\]
with some random $\a\in\C$ and $\Re\a>0$. Using \er{3dmon}
facilitates the calculations significantly because it removes a need
to solve the Riccati equation \er{ricatti} for finding the
projectors.

\section{Conclusion}

We have studied the frequency versus horizontal wavenumber spectrum
$\omega( k_{1})$ of subsonic and supersonic SH surface waves in 2D
semi-infinite phononic crystals periodic along the surface and depth
coordinates $x_{1}$ and $x_{2}.$ For instance, these may be periodic
structures of homogeneous bars parallel to the free surface
$x_{2}=0$. The necessary condition for the existence of SH surface
waves is vertical asymmetry (in $x_{2}$) of the unit-cell profile.
By analogy with the 1D-periodic case of layers parallel to the
surface (see \cite{SPG}), the subsonic SH surface waves are more
probable if the unit cell is 'slower' on the top than on the bottom.
The onset slope of this branch, i.e. the effective (quasistatic)
speed of SH surface waves, is the slope of projection on the plane
$(\omega ,k_{1})$ of the lower bound of Floquet spectrum $\omega(
\mathbf{k})$ of the infinite 2D-periodic medium. Enhancing the
unit-cell horizontal asymmetry (in $x_{1}$)  increases the band gap
which separates the subsonic branch and the next surface-wave branch
at the edge of Brillouin zone. Turning the profile upside down
replaces the SH surface wave solutions by the 'non-physical' ones
(increasing into the depth).

The method proposed herein for calculating the surface wave spectra
is based on the dispersion equation expressed through the projector
on the subspace of evanescent modes. It is defined as an integral of
the resolvent of the monodromy (transfer) matrix, and the integrand
function satisfies Ricatti equation whose coefficients are members
of 1D Fourier series of material properties. Thus there is neither a
need to identify partial modal solutions of the wave equation, nor
to calculate the monodromy matrix which is prone to numerical
instability. Knowing the projector for the evanescent modes also
yields the number $N_{\mathrm{p}}$ of the propagating modes at any
given $\omega $ and $k_{1}$. Root finding of the dispersion equation
can be refined in the non-propagating domains $N_{\mathrm{p}}=0.$
The methods can be generalized for the vector waves and the 3D case.

{\it Acknowledgement.} The authors are grateful to R. Craster, M.
Deschamps and V. Pagneux for useful discussions. A.A.K. acknowledges
support from the University Bordeaux 1 through the project AP-2011.

\end{document}